\documentclass[conference,10pt]{IEEEtran}
\usepackage{cite}
\usepackage{amsmath,amssymb,amsfonts}
\usepackage{algorithm}
\usepackage{algpseudocode}
\usepackage{textcomp}
\usepackage{xcolor}
\usepackage{algorithm}
\def\BibTeX{{\rm B\kern-.05em{\sc i\kern-.025em b}\kern-.08em
    T\kern-.1667em\lower.7ex\hbox{E}\kern-.125emX}}
\AtBeginDocument{\definecolor{tmlcncolor}{cmyk}{0.93,0.59,0.15,0.02}\definecolor{NavyBlue}{RGB}{0,86,125}}

\usepackage{soul}
\usepackage{subfig}
\usepackage{multirow}  
\usepackage[nolist]{acronym}
\usepackage{pdfpages}
\usepackage{enumitem}
\usepackage{graphicx}
\usepackage{siunitx}
\usepackage{caption}
\usepackage{algorithm}
\usepackage{algpseudocode}
\usepackage{booktabs}
\usepackage{makecell}

 \usepackage{bm}

\newcommand{\figref}[1]{\figurename~{\ref{#1}}}

\newcommand{\tabref}[1]{Table~\ref{#1}}

\begin{acronym}
    \acro{QoS}{quality of service}
    \acro{5G}{fifth generation}
    \acro{6G}{sixth generation}
    \acro{ML}{machine learning}
    \acro{MAC}{medium access control}
    \acro{O-RAN}{Open Radio Access Network}
    \acro{RAN}{Radio Access Network}
    \acro{eMBB}{Enhanced Mobile Broadband}
    \acro{URLLC}{Ultra-Reliable Low Latency Communications}
    \acro{RU}{Radio Unit}
    \acro{DU}{Distributed Unit}
    \acro{CU}{Central Unit}
    \acro{CP}{control plane}
    \acro{UP}{user plane}
    \acro{MANO}{Management and Orchestration}
    \acro{NFV}{Network Functions Virtualization}
    \acro{RIC}{Radio Access Network Intelligent Controller}
    \acro{near-RT}{near-real-time}
    \acro{non-RT}{non-real-time}
    \acro{DL}{deep learning}
    \acro{DRL}{deep reinforcement learning}
    \acro{VAE}{variational autoencoder}
    \acro{UE}{User Equipment}
    \acro{PRB}{Physical Resource Block}
    \acro{SINR}{Signal to Interference \& Noise Ratio}
    \acro{ESA}{exhaustive search algorithm}
    \acro{DNN}{deep neural network}
    \acro{MDP}{Markov Decision Process}
    \acro{DQN}{Deep Q-Network}
    \acro{BS}{Base Station}
    \acro{KL}{Kullback-Leibler}
    \acro{MSE}{mean squared error}
    \acro{MAE}{Mean Absolute Error}
	\acro{mMTC}{massive Machine-Type Communications}
	\acro{Diffusion-RL}{Diffusion-based Reinforcement Learning}
	\acro{SS-VAE}{Semi-Supervised Variational Autoencoder}
	\acro{Diffusion-QL}{Diffusion Q-Learning}
	\acro{RL}{Reinforcement Learning}
	\acro{GAN}{Generative Adversarial Network}
	\acro{BAEP}{Binary Association Error Percentage}
    \acro{TD}{Temporal Difference}
\end{acronym}

\begin{document}

\title{Generative Resource Allocation for 6G O-RAN with Diffusion Policies}

\author{
Salar Nouri$^{\dagger}$, 
Mojdeh Karbalaeimotaleb$^{\dagger}$, 
Vahid Shah-Mansouri$^{\dagger}$, 
and Tarik Taleb$^{*}$\\[3pt]
$^{\dagger}$School of Electrical and Computer Engineering, University of Tehran, Tehran, Iran\\
$^{*}$Ruhr University Bochum, Bochum, Germany\\
\texttt{\{salar.nouri, mojdeh.karbalaee, vmansouri\}@ut.ac.ir}\\
\texttt{tarik.taleb@ruhr-uni-bochum.de}
}

\maketitle

\begin{abstract}
    Dynamic resource allocation in O-RAN is critical for managing the conflicting QoS requirements of 6G network slices. Conventional reinforcement learning agents often fail in this domain, as their unimodal policy structures cannot model the multi-modal nature of optimal allocation strategies.
    This paper introduces Diffusion Q-Learning (Diffusion-QL), a novel framework that represents the policy as a conditional diffusion model. Our approach generates resource allocation actions by iteratively reversing a noising process, with each step guided by the gradient of a learned Q-function. This method enables the policy to learn and sample from the complex distribution of near-optimal actions.
    Simulations demonstrate that the Diffusion-QL approach consistently outperforms state-of-the-art DRL baselines, offering a robust solution for the intricate resource management challenges in next-generation wireless networks.
\end{abstract}

\begin{IEEEkeywords}
	Resource allocation, Network slicing, Reinforcement Learning, Diffusion Model, Generative AI
\end{IEEEkeywords}

\section{INTRODUCTION}
\label{introduction}

\IEEEPARstart{T}{he} vision for \ac{6G} wireless networks, enabled by flexible architectures like \ac{O-RAN}, promises transformative applications such as holographic communications and the tactile internet \cite{wang2023road, strinati2025toward}. Achieving this requires addressing a critical resource management challenge: supporting network slices with conflicting \ac{QoS} requirements. \ac{eMBB} demands peak rates above 10 Gbps, \ac{URLLC} requires sub-millisecond latency with near-perfect reliability, and massive \ac{mMTC} must serve extremely high device densities \cite{motalleb2022resource, oranwhitepaper}. This creates a complex, high-dimensional, dynamic optimization problem beyond traditional allocation methods, motivating intelligent network control.

The \ac{DRL} has emerged as a promising paradigm for this challenge, with recent efforts exploring various architectures. Advanced methods have leveraged graph neural networks to capture topological complexities in V2X systems \cite{ji2025gnn}, multi-agent frameworks for distributed control \cite{chen2024madrl, yan2024madrl}, and federated learning to enhance privacy and scalability \cite{zhao2024feddrl}. However, a critical research gap persists: these approaches commonly rely on simple, unimodal policy parameterizations, such as Gaussian distributions. Such policies are ill-equipped to capture the complex, often multi-modal, nature of the optimal resource allocation solution space, where multiple distinct allocation strategies can yield similar performance \cite{zhu2023diffusion}. This fundamental limitation in policy expressiveness leads to suboptimal performance and poor generalization in the dynamic \ac{O-RAN} environment.

Recognizing this policy expressiveness gap, generative models have been investigated to create more powerful policies. While our prior work introducing a \ac{SS-VAE} \cite{nouri2024semi} demonstrated the potential of this direction, VAEs can suffer from training instability and a less expressive latent space. To truly solve this problem, a more robust framework is needed. This paper bridges this gap by introducing a novel \ac{DRL} framework, \textbf{Diffusion-QL}, for joint resource allocation and network slicing in \ac{O-RAN}. We directly confront the limitations of prior methods by leveraging a conditional diffusion model—a powerful class of generative models renowned for its stability and ability to learn complex distributions \cite{wang2022diffusion}. We formulate the slicing problem for three distinct services—\ac{eMBB}, \ac{URLLC}, and \ac{mMTC}—and deploy our \ac{Diffusion-QL} agent as an xApp within the near-real-time \ac{RIC}. Unlike conventional methods that learn a single action, our framework generates a distribution of near-optimal actions guided by a learned Q-function, enabling more expressive and adaptive decision-making.

The main contributions of this paper are summarized as follows:
\begin{itemize}
    \item \textit{O-RAN Slicing Model:} We formally model the joint power and \ac{PRB} allocation problem for three key \ac{O-RAN} services: \ac{eMBB}, \ac{URLLC}, and \ac{mMTC}, capturing their diverse \ac{QoS} requirements.
    
    \item \textit{Diffusion-QL Framework:} We design and develop \ac{Diffusion-QL}, a generative \ac{DRL} agent that uses a Q-guided diffusion model to overcome the policy expressiveness limitations of prior \ac{DRL} approaches.
    
    \item \textit{Performance and Robustness Validation:} Through extensive system-level simulations, we demonstrate that \ac{Diffusion-QL} significantly outperforms state-of-the-art baselines. We further validate its \textit{robustness and reliable performance} across diverse scenarios, showcasing its better adaptability.
    
    \item \textit{Theoretical Complexity Analysis:} We provide a theoretical computational analysis of our framework, confirming its feasibility for real-time operation and decision-making within the stringent latency constraints of the near-RT \ac{RIC}.
\end{itemize}

\section{System Model and Problem Formulation}
\label{system_model}

\begin{figure}[t]
	\centerline{\includegraphics[width=0.45\textwidth]{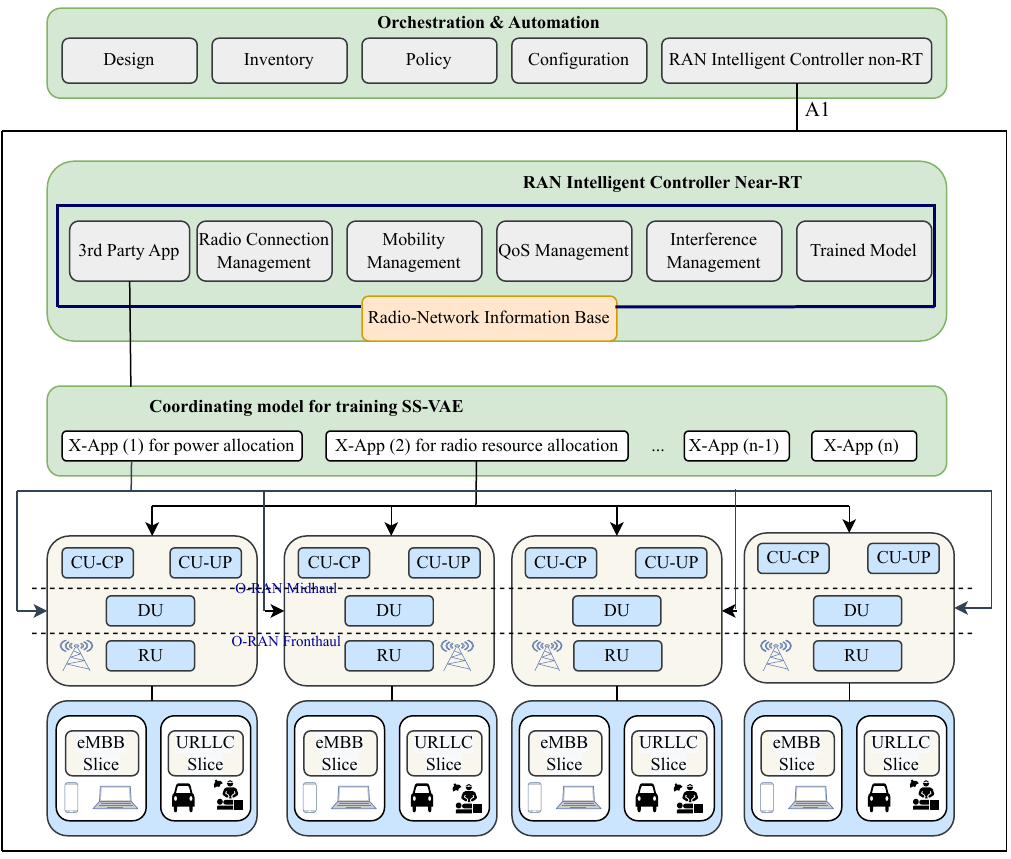}}
	\caption{Architectural overview of the O-RAN system\cite{nouri2024semi}.}
	\label{fig:oran system model}
\end{figure}

\subsection{System Model}
In this study, we address the joint allocation of transmission power and \acp{PRB} within an \ac{O-RAN} architecture to support three primary network slices: \ac{eMBB}, \ac{URLLC}, and \ac{mMTC}, each with distinct \ac{QoS} requirements for latency and throughput. Our model assumes each \ac{RU} serves multiple \acp{UE} distributed across these slices. We consider $S_1$, $S_2$, and $S_3$ slices for \ac{eMBB}, \ac{URLLC}, and \ac{mMTC}, respectively. Each service type $j \in \{e,u,m\}$ comprises $S_j$ slices, and each slice $s_j$ serves $U_{s_j}$ \acp{UE}. Consistent with \cite{nouri2024semi}, we jointly optimize power and radio resources, assuming single-antenna \acp{UE} and \acp{RU} (\figref{fig:oran system model}).

The system's spectrum is divided into $K$ \acp{PRB}, shared among $R$ single-antenna \acp{RU}. Each slice $s$ is allocated $\bar{K}_s$ PRBs, satisfying $\sum_s \bar{K}_s \le K$. The binary variable $\alpha_{u_s}^r=1$ indicates the association of \ac{UE} $u$ in slice $s$ to \ac{RU} $r$, where each \ac{UE} connects to exactly one \ac{RU} which is shown as $\sum_{r} \alpha_{u_s}^r = 1, \quad \forall u, s.$
The \ac{PRB} allocation is indicated by the binary variable $\beta_{u_s,k}^r=1$ if PRB $k$ of \ac{RU} $r$ is assigned to \ac{UE} $u$ in slice $s$. This assignment is valid only if the \ac{UE} is associated with the \ac{RU}, and each \ac{PRB} is exclusively allocated to at most one \ac{UE} per \ac{RU}, as enforced by:
\begin{align}
	\beta_{u_s,k}^r \leq \alpha_{u_s}^r, \; \; \; \; \forall r \in \mathcal{R}, \; \; k \in \mathcal{K_s}, \; \; u \in \mathcal{U}, \; \; s \in \mathcal{S}, \\
	\sum_{u \in \mathcal{U}} \alpha_{u_s}^r \beta_{u_s,k}^r \leq 1,   \; \; \; \; \forall r \in \mathcal{R}, k \in \mathcal{K_s}, \; \; s \in \mathcal{S}.
\end{align}
This exclusive allocation per \ac{RU} eliminates intra-cell interference, though inter-cell interference is considered in our system model.
The \ac{SINR} of the $u^{\mathrm{th}}$ UE in slice $s$ served by O-RU $r$ on PRB $k$ is given by
\begin{equation}\label{eq1}
    \rho_{u_s,k}^r = \frac{\alpha_{u_s}^r \beta_{u_s,k}^r p_{u_s,k}^r|h_{u_s,k}^r|^2}{BN_0 + I_{u_s,k}^r},
\end{equation}
where $I_{u_s,k}^r$ represents the inter-cell interference, defined as
\begin{equation}\label{eq2}
    I_{u_s,k}^r = {\sum_{j \neq r}^{R} \sum_{l=1, l \neq s}^{S} \sum_{i=1, i \neq u}^{U} \alpha_{i_l}^j \beta_{i_l,k}^j p_{i_l,k}^j |h_{i_l,k}^j|^2}.
\end{equation}
In addition, $|h_{u_s,k}^r|^2$ is the channel power gain between \ac{UE} $u$ in slice $s$ and O-\ac{RU} $r$ on \ac{PRB} $k$, and $p_{u_s,k}^r$ is the transmission power allocated to \ac{UE} $u$ in slice $s$ by O-\ac{RU} $r$ on \ac{PRB} $k$. $B$ is the bandwidth, and $N_0$ is the Gaussian noise power spectral density, so $BN_0$ is the noise power of the system.
The achievable data rate of the $u^{\mathrm{th}}$ \ac{UE} in slice $s$ served by O-\ac{RU} $r$ on \ac{PRB} $k$ is given by $R_{u_s,k}^r = \alpha_{u_s}^r \beta_{u_s,k}^r B
\log_2\!\left(1 + \rho_{u_s,k}^r\right),$
where $B$ denotes the \ac{PRB} bandwidth. The total achievable rate of \ac{UE} $u$ in slice $s$ is $    R_{u_s} = \sum_{r \in \mathcal{R}} \sum_{k \in \mathcal{K}_s} 
    \alpha_{u_s}^r \beta_{u_s,k}^r B 
    \log_2\!\left(1 + \rho_{u_s,k}^r\right).$
Accordingly, the total throughput of slice $s$ and the overall system throughput are, respectively,
$R_s = \sum_{u \in \mathcal{U}_s} R_{u_s}$ and 
$R_{\mathrm{tot}} = \sum_{s \in \mathcal{S}} R_s$.
For simplicity, we model the transmission delay $D_{u_s}$ for a packet of average size $L_{u_s}$ as a function of the achievable rate $R_{u_s}$ which is $D_{u_s} = \frac{L_{u_s}}{R_{u_s}}.$
The instantaneous capacity of \ac{RU} $r$, denoted $C_r$, is the aggregate data rate of all \acp{UE} it serves, given by $C_r = \sum_{s \in \mathcal{S}} \sum_{u \in \mathcal{U}_s} \sum_{k \in \mathcal{K}_s} 
\alpha_{u_s}^r \beta_{u_s,k}^r B 
\log_2\!\left(1 + \rho_{u_s,k}^r\right).$
We model the transmission delay $D_{u_s}$ for a packet of average size $L_{u_s}$ bits, ignoring queuing and processing delays for simplicity, as $D_{u_s} = \frac{L_{u_s}}{R_{u_s}}.$
The instantaneous capacity of O-\ac{RU} $r$, denoted $C_r$, is the aggregate data rate of all \acp{UE} it serves, given by $C_r = \sum_{s \in \mathcal{S}} \sum_{u \in \mathcal{U}_s} \sum_{k \in \mathcal{K}_s} 
\alpha_{u_s}^r \beta_{u_s,k}^r B 
\log_2\!\left(1 + \rho_{u_s,k}^r\right).$
\subsection{Problem Formulation}
Given that the joint resource allocation problem is NP-hard, we formulate an optimization to find a near-optimal solution. The objective is to maximize the total achievable rate of all slices (Equation \eqref{eq:opt_obj}), subject to constraints on delay, power, and O-RU capacity (Equations \eqref{eq:delay_cons} -- \eqref{eq:capacity_cons}). The problem is formulated as:
\allowdisplaybreaks
\begin{align}
\max_{\{\alpha,\beta,p\}} \quad 
& \sum_{s \in \mathcal{S}} \sum_{u \in \mathcal{U}_s} \sum_{r \in \mathcal{R}} \sum_{k \in \mathcal{K}_s}
\alpha_{u_s}^r \beta_{u_s,k}^r B
\log_2\!\left(1 + \rho_{u_s,k}^r\right), \label{eq:opt_obj} \\
\text{s.t.} \quad 
& D_{u_s} = \frac{L_{u_s}}{R_{u_s}} \le D_{u_s}^{\max}, 
\quad \forall u,s, \label{eq:delay_cons} \\
& \sum_{s \in \mathcal{S}} \sum_{u \in \mathcal{U}_s} \sum_{k \in \mathcal{K}_s}
\alpha_{u_s}^r \beta_{u_s,k}^r p_{u_s,k}^r \le P_r^{\max}, 
\quad \forall r, \label{eq:power_cons} \\
& C_r \le C_r^{\max}, 
\quad \forall r, \label{eq:capacity_cons} \\
& \sum_{r} \alpha_{u_s}^r = 1, 
\quad \forall u,s, \\
& \sum_{u} \alpha_{u_s}^r \beta_{u_s,k}^r \le 1,
\quad \forall r,k,s, \\
& \sum_{s} \bar{K}_s \le K, \quad
\alpha_{u_s}^r, \beta_{u_s,k}^r \in \{0,1\}.
\end{align}

\section{Methodology}
\label{Proposed_scheme}

\begin{figure}[t]
	\centering
    {\includegraphics[width=0.4\textwidth]{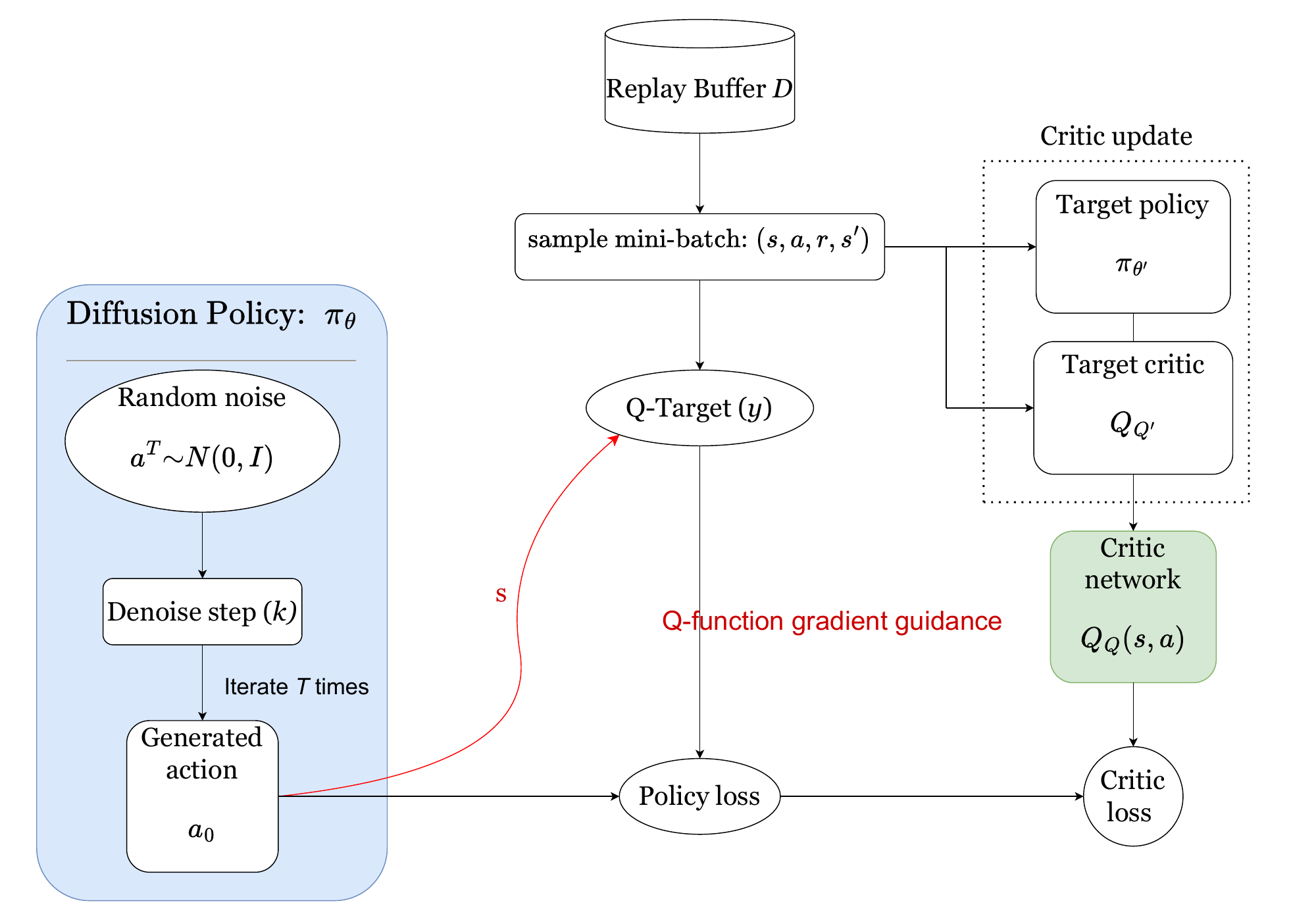}}
	\caption{Overview of the Diffusion-QL training loop.}
	\label{fig: model_overview_diffusion-ql}
\end{figure}

This study proposes a \ac{Diffusion-RL} model for resource allocation in the \ac{O-RAN} architecture, chosen for its ability to efficiently explore high-dimensional state spaces while providing better generalization and robustness over traditional \ac{RL} methods \cite{zhu2023diffusion}. These characteristics are critical for managing the dynamic and uncertain conditions of \ac{6G} networks. By integrating a generative model as the policy, our approach achieves a more effective exploration-exploitation balance, handling data sparsity and mitigating the high computational costs associated with conventional online \ac{RL} training. The overview of our proposed methodology is illustrated in \figref{fig: model_overview_diffusion-ql}.

To validate the performance of our model, we compare it against three key benchmarks detailed in \cite{nouri2024semi}. The \ac{ESA} provides a theoretical optimum but is computationally infeasible for real-time deployment. Our prior work, the \ac{SS-VAE}, also uses a generative model but requires a labeled dataset generated by the \ac{ESA}. Finally, the \ac{DQN} serves as a standard \ac{DRL} baseline but is known to be resource-intensive and can struggle with generalization.

\subsection{\ac{Diffusion-RL}}
\label{sec:diffusion_rl}

To address the complex resource allocation problem in \ac{O-RAN}, we propose \ac{Diffusion-QL}, a novel framework that represents the \ac{RL} policy as a conditional diffusion model. This approach overcomes the limitations of conventional policies (e.g., Gaussian), which are unimodal and fail to capture the multi-modal nature of optimal allocation strategies \cite{wang2022diffusion}. Diffusion models are highly expressive generative models that excel at learning complex distributions with superior training stability compared to alternatives like \acp{VAE} \cite{ho2020denoising, zhu2023diffusion}.

Our \ac{Diffusion-QL} policy learns to generate actions by iteratively reversing a gradual noising process. The training objective is twofold: it implicitly performs behavior cloning by learning to denoise samples, ensuring generated actions remain close to the data distribution, while simultaneously being guided by the gradient of a learned Q-function to maximize long-term returns.

The policy, $\pi_\theta(a|s)$, is formally represented by the reverse process of a conditional diffusion model. Let $a_0$ be an action vector. The forward process $q$ is a fixed Markov chain that gradually adds Gaussian noise over $T$ steps according to a variance schedule $\{\beta_t\}_{t=1}^T$. The distribution of a noised action $a_t$ at any step $t$ can be sampled directly from the original action $a_0$ \cite{ho2020denoising} which is shown as $q(a_t | a_0) = \mathcal{N}(a_t; \sqrt{\bar{\alpha}_t} a_0, (1 - \bar{\alpha}_t) \mathbf{I}),$
where $\alpha_t = 1 - \beta_t$ and $\bar{\alpha}_t = \prod_{i=1}^t \alpha_i$.

The reverse process, which constitutes the policy, is parameterized by a neural network $\epsilon_\theta(a_t, t, s)$ trained to predict the noise $\epsilon$ added at step $t$, conditioned on the network state $s$. The network is optimized by minimizing the denoising score matching loss, which serves as an implicit behavior cloning objective \cite{wang2022diffusion} that is shown as
$\mathcal{L}_{diffusion}(\theta) = \mathbb{E}_{t, s, a_0, \epsilon} \left[ ||\epsilon - \epsilon_\theta(\sqrt{\bar{\alpha}_t}a_0 + \sqrt{1-\bar{\alpha}_t}\epsilon, t, s) ||^2 \right].$
To guide the policy towards high-value actions, we integrate a critic network, $Q_\phi(s, a)$, which is trained using the standard \ac{TD} learning objective from a replay buffer $\mathcal{D}$ \cite{wang2022diffusion, zhu2023diffusion} which is represented as $\mathcal{L}_{critic}(\phi) = \mathbb{E}_{(s, a, r, s') \sim \mathcal{D}} \left[ \left( Q_\phi(s, a) - (r + \gamma Q_{\phi_{tgt}}(s', a')) \right)^2 \right].$

During the reverse denoising process, the gradient of this Q-function is used to perturb the sampling steps. This explicitly steers the generated action towards regions of the action space with higher estimated long-term returns. The mean of the distribution for the next denoising step, $a_{t-1}$, is adjusted as  $\hat{\mu}_\theta(a_t, t, s) = \mu_\theta(a_t, t, s) + w \cdot \Sigma_t \nabla_{a_t} Q_\phi(s, a_t)$ \cite{wang2022diffusion}.
where $\mu_\theta$ is the standard predicted mean from the denoising network, $\Sigma_t$ is the reverse step covariance, and $w$ is a guidance scale hyperparameter. This integration of the critic's gradient forms a cohesive \ac{Diffusion-QL} algorithm that balances exploration (via the diffusion model's generative nature) and exploitation (via Q-function guidance). The training procedure for our \ac{Diffusion-QL} algorithm is summarized in Algorithm \ref{alg:diffusion_ql}.

\begin{algorithm}[t]
\caption{\ac{Diffusion-QL} for O-RAN Resource Allocation}
\label{alg:diffusion_ql}
\begin{algorithmic}[1]
\State \textbf{Initialize:} Diffusion policy $\pi_\theta$, critic networks $Q_{\phi_1}, Q_{\phi_2}$.
\State \textbf{Initialize:} Target networks $\pi_{\theta'} \leftarrow \pi_\theta$, $Q_{\phi'_1} \leftarrow Q_{\phi_1}$, $Q_{\phi'_2} \leftarrow Q_{\phi_2}$.
\State \textbf{Initialize:} Replay buffer $\mathcal{D}$.
\For{each training iteration}
    \State Sample a mini-batch $B = \{(s, a, r, s')\} \sim \mathcal{D}$.
    \State
    \State \textit{// Critic Update}
    \State Sample next action $a' \sim \pi_{\theta'}(\cdot | s')$ via the reverse diffusion process.
    \State Compute target Q-value:
    \State $y \leftarrow r + \gamma \min_{i=1,2} Q_{\phi'_i}(s', a')$
    \State Update critics $\phi_1, \phi_2$ by minimizing:
    \State $\mathcal{L}_{critic}(\phi_i) = \mathbb{E}_{(s,a) \in B} \left[ (Q_{\phi_i}(s,a) - y)^2 \right]$
    \State
    \State \textit{// Policy Update}
    \State Sample action $a_0 \sim \pi_\theta(\cdot | s)$ via the reverse diffusion process.
    \State Compute the combined policy loss:
    \State $\mathcal{L}_{policy} = \mathcal{L}_{diffusion}(\theta) - \lambda \cdot \mathbb{E}_{s \in B, a^0 \sim \pi_\theta}[Q_{\phi_1}(s, a^0)]$
    \State Update policy $\theta$ by minimizing $\mathcal{L}_{policy}$.
    \State
    \State \textit{// Target Network Update}
    \State $\theta' \leftarrow \rho\theta' + (1-\rho)\theta$
    \State $\phi'_i \leftarrow \rho\phi'_i + (1-\rho)\phi_i$ for $i \in \{1,2\}$
\EndFor
\end{algorithmic}
\end{algorithm}

\subsection{\ac{MDP} Formulation for \ac{Diffusion-QL}}
We formulate the resource allocation problem as an \ac{MDP}, enabling the \ac{O-RAN} orchestrator to act as a learning agent, as follows:

\textbf{State:} The state at time $t$, $\mathfrak{s}(t) \in \mathfrak{S}$, is represented by $\{\mathfrak{s}_{u}(t)\}_{u=1}^{U}$, where $\mathfrak{s}_{u}(t)$ is a binary indicator that equals one if the data rate requirement of \ac{UE} $u$ is satisfied and zero otherwise. The state also corresponds to a quantized transmission power level, providing a snapshot of both \ac{QoS} satisfaction and current resource distribution.

\textbf{Action:} An action $\mathfrak{a} \in \mathfrak{A}$ represents the complete resource allocation decision for all \acp{UE}. It is a composite vector containing the \ac{UE}-\ac{RU} association indicators and the \ac{PRB} assignments: $\mathfrak{a} = \{\alpha_{u,b,s}, \{\beta_{u,b,m,s}\}_{m=1}^{M}\}_{b=1}^{B}$.

\textbf{Reward:} 
The reward function $\mathfrak{R}(\mathfrak{s}, \mathfrak{a})$, which evaluates the quality of a given state-action pair, is a weighted combination of the primary data rate objective and terms representing system constraints as $\mathfrak{R}(\mathfrak{s}, \mathfrak{a}) = \Theta_{r}R_{u_{s}} + \Theta_{\text{const}}C_{u_{s},m}^{b} + \Theta_{\text{bias}},$
where $\Theta_{r}$, $\Theta_{\text{const}}$, and $\Theta_{\text{bias}}$ are the respective weights assigned to the data rate, the constraints, and a bias value. This structure guides the agent's learning process by indicating the desirability of various actions across different network states. 
\section{SIMULATION RESULTS}
\label{simulation_results}
\begin{table}[t]
    \centering
    \caption{Simulation Environment and Hyperparameter Settings.}
    \label{tab:simulation_parameters}
    \begin{tabular}{l c}
        \toprule
        \textbf{Parameter} & \textbf{Value} \\
        \midrule
        \multicolumn{2}{l}{\textit{\textbf{O-RAN Environment}}} \\
        Cell Radius & 400 m \\
        Number of PRBs & 50 \\
        gNB Transmit Power, $P_{max}$ & 46 dBm \\
        Noise Power Spectral Density & -174 dBm/Hz \\
        Channel Model & 3GPP TR 38.901 \\
        Path Loss Model & Urban Macro (path loss exponent: 3.76) \\
        User Distribution & 40/40/20\% (eMBB/URLLC/mMTC) \\
        \midrule
        \multicolumn{2}{l}{\textit{\textbf{QoS Requirements}}} \\
        eMBB Min. Rate Req. & 10 Mbps \\
        URLLC Min. Rate Req. & 2 Mbps \\
        URLLC Max. Delay Req. & 1 ms \\
        \midrule
        \multicolumn{2}{l}{\textit{\textbf{RL Training}}} \\
        Optimizer & Adam \\
        Critic Learning Rate & 3e-4 \\
        Policy Learning Rate & 1e-4 \\
        Discount Factor, $\gamma$ & 0.98 \\
        Target Network Update Rate, $\rho$ & 0.005 \\
        Replay Buffer Size & 200,000 \\
        Batch Size & 128 \\
        \midrule
        \multicolumn{2}{l}{\textit{\textbf{Diffusion-QL Model}}} \\
        Network Architecture & 3-Layer MLP \\
        Neurons per Layer & 128 \\
        Diffusion Timesteps, $T$ & 20 \\
        Guidance Scale, $w$ & 1.2 \\
        Noise Schedule, $\beta_t$ & Linear, 1e-4 to 2e-2 \\
        \bottomrule
    \end{tabular}
\end{table}

\begin{table}[t]
	\centering
	\caption{Generalization performance against the ESA benchmark on test data.}
	\label{tab:perf_metrics}
	\begin{tabular}{l c c c c}
		\toprule
		\textbf{Algorithm} & \textbf{MAE} $\downarrow$ & \textbf{$R^2$} $\uparrow$ & \textbf{Cosine Sim.} $\uparrow$ & \textbf{BAEP (\%)} $\downarrow$ \\
		\midrule
		\ac{SS-VAE} & 0.1407 & 0.7237 & 0.9745 & 5.45 \\
		\ac{DQN} & 0.2156 & 0.6985 & 0.8915 & 9.72 \\
		\ac{Diffusion-QL} & \textbf{0.1381} & \textbf{0.7461} & \textbf{0.9808} & \textbf{4.19} \\
		\bottomrule
	\end{tabular}
\end{table}

\begin{figure*}[t]
\scriptsize
	\centering
	\subfloat[Reward Convergence.]{
		\includegraphics[width=0.32\linewidth]{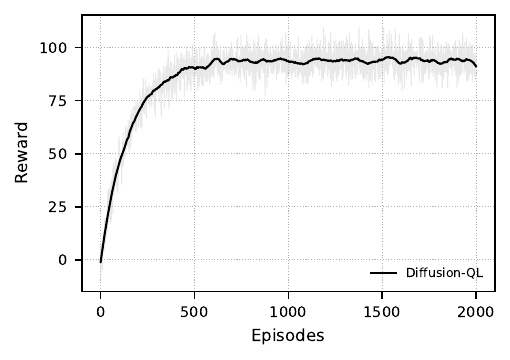}
		\label{fig:reward_convergence}
	} \hfill
	\subfloat[Throughput vs. Number of UEs.]{
		\includegraphics[width=0.32\linewidth]{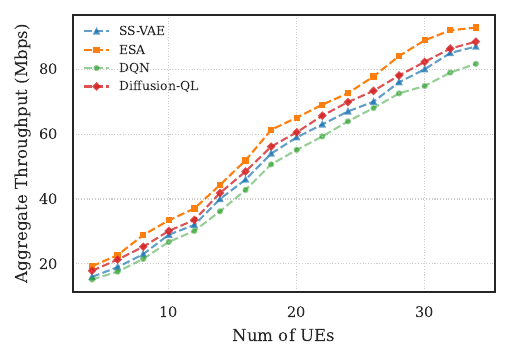}
		\label{fig:agg_thp_algorithms}
	} \hfill
	\subfloat[Throughput Sensitivity (Power \& UEs).]{
		\includegraphics[width=0.32\linewidth]{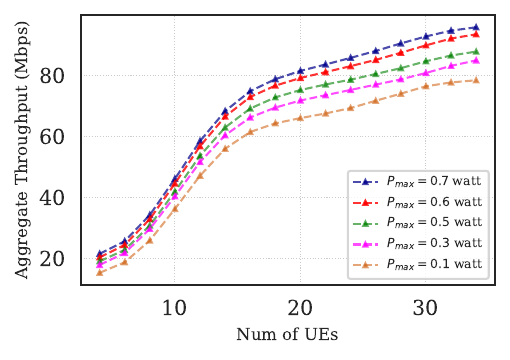}
		\label{fig:agg_thr_diff-ql_p-ru}
	} \\
	\subfloat[Per-Slice Throughput vs. RU Power.]{
		\includegraphics[width=0.32\linewidth]{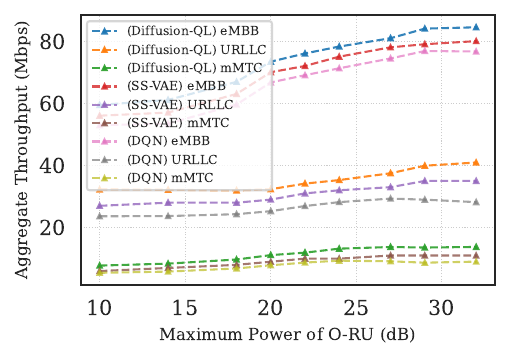}
		\label{fig:agg_throughput_ru_power_all_alg}
	} \hfill
	\subfloat[Per-Slice Throughput vs. Slice Power.]{
		\includegraphics[width=0.32\linewidth]{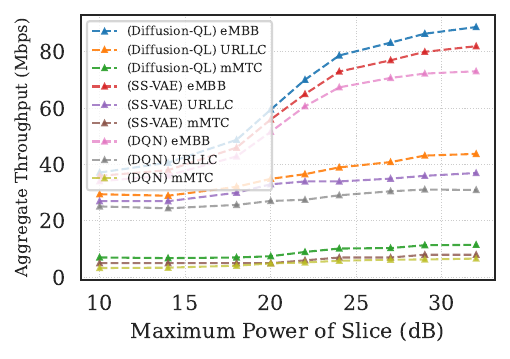}
		\label{fig:agg_throughput_slice_power_all_alg}
	} \hfill
	\subfloat[Robustness to Interference.]{
		\includegraphics[width=0.32\linewidth]{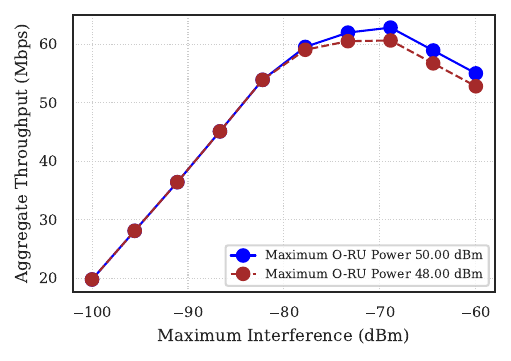}
		\label{fig:agg_throughput_vs_interference}
	}	
	\caption{Performance evaluation of Diffusion-QL against benchmarks. (a) Reward convergence during training. (b) Aggregate throughput scalability with increasing UEs. (c) Throughput sensitivity to RU power. (d, e) Per-slice throughput analysis under varying power constraints. (f) Robustness to inter-cell interference}
	\label{fig:performance_of_algorithms}
\end{figure*}

This section provides an evaluation of our proposed \ac{Diffusion-QL} algorithm against three benchmarks: \ac{ESA}, \ac{DQN}, and \ac{SS-VAE} \cite{nouri2024semi}. All experiments were conducted in PyTorch \cite{paszke2019pytorch} on a workstation equipped with an NVIDIA Volta V100 GPU. The detailed parameters for the \ac{O-RAN} simulation environment and our \ac{Diffusion-QL} model are summarized in \tabref{tab:simulation_parameters}. To ensure a fair comparison, the configurations for all benchmark algorithms are identical to those used in \cite{nouri2024semi}.

\subsection{Training Stability and Generalization Accuracy}
We first analyze the learning dynamics and generalization capability of the \ac{Diffusion-QL} agent. \figref{fig:reward_convergence} plots the average reward per episode over the course of training. The curve shows a rapid increase in performance within the first 500 episodes, followed by stable convergence to a high reward value. This demonstrates that the agent not only learns an effective resource allocation policy quickly but also maintains stable performance without divergence, a critical requirement for reliable \ac{DRL}-based network controllers.

To quantify the model's ability to generalize and produce near-optimal allocation decisions, we evaluated it on a held-out test dataset. \tabref{tab:perf_metrics} compares the performance of the learning-based methods against the optimal solutions found by the \ac{ESA}. Our \ac{Diffusion-QL} model achieves the best results across all four metrics: it has the lowest \ac{MAE} and \ac{BAEP}, and the highest $R^2$ Score and Cosine Similarity. The \ac{BAEP}, which measures the error in the binary allocation decisions (\ac{UE} association and \ac{PRB} assignment), is particularly telling. With a \ac{BAEP} of only 4.19\%, \ac{Diffusion-QL} demonstrates a superior ability to replicate the discrete allocation structure of the optimal policy compared to both \ac{SS-VAE} (5.45\%) and \ac{DQN} (9.72\%). This confirms that the expressive, generative nature of the diffusion policy allows it to learn a more accurate and generalizable representation of the optimal solution.

\subsection{Scalability and Throughput Performance}
A critical scenario for any resource allocation algorithm is its ability to scale with increasing network load. \figref{fig:agg_thp_algorithms} evaluates this by plotting the aggregate system throughput as the number of \acp{UE} increases from 5 to 35. In this scenario, \ac{Diffusion-QL} consistently outperforms the other learning-based methods, \ac{SS-VAE} and \ac{DQN}, and closely tracks the performance of the computationally infeasible \ac{ESA}. This demonstrates that our approach finds near-optimal solutions across a wide range of user densities. For all algorithms, the throughput begins to saturate around 30-35 \acp{UE}, which is expected as the system becomes resource-limited by the fixed number of \acp{PRB} and the maximum transmit power.

\figref{fig:agg_thr_diff-ql_p-ru} further explores the behavior of our \ac{Diffusion-QL} model under varying power constraints. As expected, increasing the maximum transmit power of the O-\ac{RU} consistently yields higher aggregate throughput across all \ac{UE} densities. This is because higher power improves the \ac{SINR} for all users, enabling higher-order modulation and coding schemes. This result serves as a validation, confirming that our agent's learned policy correctly responds to fundamental changes in the physical layer environment.

\subsection{Per-Slice Performance and Power Sensitivity}
Beyond aggregate performance, it is crucial to evaluate how well each algorithm manages the conflicting requirements of different network slices. \figref{fig:agg_throughput_ru_power_all_alg} and \figref{fig:agg_throughput_slice_power_all_alg} provide this insight by showing the per-slice throughput as a function of maximum O-\ac{RU} power and maximum slice-specific power, respectively.

In both scenarios, \ac{Diffusion-QL} achieves the highest throughput for every single slice—\ac{eMBB}, \ac{URLLC}, and \ac{mMTC}—when compared to the \ac{SS-VAE} and \ac{DQN} baselines. This is a significant finding, as it demonstrates that the performance gains are not achieved by sacrificing one service for another. Instead, \ac{Diffusion-QL} finds a more globally efficient allocation that benefits all slices simultaneously. As expected, the high-data-rate \ac{eMBB} slice achieves the highest throughput, followed by \ac{URLLC} and \ac{mMTC}, confirming that the agent correctly prioritizes resources according to the service requirements.

\subsection{Robustness to Inter-Cell Interference}
Finally, we test the robustness of our algorithm in a more challenging, interference-limited scenario. \figref{fig:agg_throughput_vs_interference} illustrates the impact of increasing inter-cell interference on the aggregate throughput of the \ac{Diffusion-QL} agent. The x-axis represents the maximum power of an external interference source. The plot reveals an inverted U-shaped curve: throughput initially peaks and then degrades as the interference becomes strong enough to lower the \ac{SINR} across the cell significantly. This behavior is expected in an interference-limited system.

Crucially, the figure shows that operating with a higher RU transmit power (50 dBm vs. 48 dBm) provides a consistent throughput advantage and pushes the point of performance degradation further to the right. This demonstrates that the learned policy is robust and can leverage available power to effectively mitigate external interference, a key capability for deployment in dense, real-world cellular networks.

\subsection{Computational Complexity Analysis}
\label{sec:complexity}

\begin{table}[t]
	\centering
	\caption{Computational Complexity for Each Algorithm.}
	\label{table:computational_costs}
	\renewcommand{\theadfont}{\bfseries}
	\begin{tabular}{l l}
		\toprule
		\thead{Method} & \thead{Computational Complexity} \\
		\midrule
		\ac{ESA} & \makecell[l]{$O(|U| \cdot |B| \cdot (M+1)! \cdot P_{\text{levels}})$} \\
		\addlinespace
		\ac{DQN} & $O(E \cdot T \cdot (|S| \cdot |A| + F))$ \\
		\addlinespace
		\ac{SS-VAE} & $O(E \cdot (D \cdot L + L^2)) + O(N \cdot D \cdot L)$ \\
		\addlinespace
		\ac{Diffusion-QL} & \makecell[l]{$O(E \cdot T \cdot (|S| \cdot |A| + F_{\text{diff}}))$ \\ where $F_{\text{diff}} = O(N_{\text{steps}} \cdot N_{\text{neurons}})$} \\
		\bottomrule
	\end{tabular}
\end{table}

To provide a complete performance comparison, we analyze the computational complexity of each algorithm. The theoretical costs for a single training run are summarized in \tabref{table:computational_costs}, where the parameters are defined as follows: $E$ is the number of episodes (or epochs), $T$ is the steps per episode, $|S|$ and $|A|$ are the state and action space sizes, $F$ is the complexity of a neural network pass, $|U|$ is the number of \acp{UE}, $|B|$ is the number of \acp{RU}, $M$ is the number of \acp{PRB}, $P_{\text{levels}}$ is the number of discrete power levels, $D$ and $L$ are the input and latent dimensions for the VAE, $N$ is the number of samples, and $N_{\text{steps}}$ is the number of diffusion denoising steps.

The cost for each algorithm is derived from its core operational loop. The \ac{ESA} exhibits factorial complexity, $O(|U| \cdot |B| \cdot (M+1)! \cdot P_{\text{levels}})$, as it must enumerate every possible combination of UE-RU association, \ac{PRB} assignment, and power level, making it computationally infeasible for any non-trivial network. The learning-based methods all have polynomial complexity. The cost for \ac{DQN} and \ac{Diffusion-QL} is dominated by the training loop over episodes and steps, where each step involves environment interaction and a network update. The complexity of \ac{SS-VAE} is determined by its training epochs over the dataset, as detailed in \cite{nouri2024semi}.

A key distinction lies in the complexity of the forward pass, $F$. For \ac{DQN} and \ac{SS-VAE}, this is a single pass through a standard neural network. For our \ac{Diffusion-QL} model, the effective forward pass, $F_{\text{diff}}$, is significantly more complex as it requires $N_{\text{steps}}$ iterations of the denoising network to generate a single action. This leads to a critical trade-off: while \ac{Diffusion-QL} achieves better performance due to its expressive generative policy, it incurs a higher computational cost during inference. This increased latency at decision-making time is a known characteristic of diffusion models and a primary consideration for deployment in real-time control loops.

\section{CONCLUSION}
\label{conclusions}

This paper introduced \ac{Diffusion-QL}, a novel framework for dynamic resource allocation in \ac{O-RAN} that leverages a Q-guided diffusion model as a highly expressive policy. Our work addresses the NP-hard problem of jointly allocating power and \acp{PRB} to satisfy the conflicting \ac{QoS} demands of \ac{eMBB}, \ac{URLLC}, and \ac{mMTC} services. Simulation results demonstrated that \ac{Diffusion-QL} significantly outperforms state-of-the-art \ac{DRL} methods, including \ac{DQN} and \ac{SS-VAE} models, particularly in satisfying stringent \ac{URLLC} constraints while maintaining competitive network throughput.

The success of our approach stems from the diffusion policy's ability to model the complex, multi-modal distribution of optimal allocation strategies—a critical limitation of conventional \ac{DRL} agents with unimodal policies. By generatively constructing actions, \ac{Diffusion-QL} achieves superior exploration and robustness in dynamic \ac{O-RAN} environments. Unlike the computationally prohibitive \ac{ESA} or the label-dependent \ac{SS-VAE}, our method provides a scalable and flexible solution that learns effectively without requiring pre-generated optimal datasets, making it highly suitable for real-world deployment. Future work will focus on exploring advanced sampling techniques to accelerate the iterative denoising process, addressing the computational demands of deploying diffusion policies in the latency-critical control loops of the \ac{O-RAN} architecture.

\bibliographystyle{IEEEtran}
\small{\bibliography{ref.bib}}

\vfill\pagebreak

\end{document}